\documentstyle [12pt]{article}
\def\be{\begin{equation}}
\def\ee{\end{equation}}

\def\cH{{\cal H}}
\def\tcH{\widetilde{\cal H}}
\def\C+{{\bf C}_+}
\def\Z{{\bf Z}}

\def\cF{{\cal F}}
\def\tD{\widetilde{D}}
\def\tL{\widetilde{L}}
\def\tV{\widetilde{V}}

\begin{document}

\begin{center}
{\Huge \bf
One-Dimensional Exactly Solvable\\
\vskip 0.6cm
Model of Polaron }

\vskip 2cm

{\large
Yu.A.Kuperin$^1 $, B.S.Pavlov$^{1,2}$, R.V.Romanov$^1$ and
G.E.Rudin$^1$
}
\vskip 8mm

$^1$ Laboratory of Quantum Networks, \\ Institute for
Physics, \\St.Petersburg State University \\198504 St.Petersburg,
Russia\\
E-mail: kuperin@JK1454.spb.edu
\vskip 0.4cm
$^2$ Department of Mathematics,
University of Auckland, Auckland, New Zealand
\vskip 0.4cm
\end{center}
\vskip 1cm

\vskip 4mm
\begin{center}
{\large Abstract}
\end{center}

Using for unperturbed electron and phonon Hamiltonians a
representation by the Jacobi matrices a one-dimensional model of
the electron-phonon interaction is constructed. In frame of the
model the polaron and scattering spectral bands are calculated
explicitly as well as the corresponding states. An explicit formula
for the polaron effective mass is derived and the fit of model
parameters is suggested. A connection of the model with the
description of phonon degrees of freedom for quantum wires is
discussed.

\newpage

\section{Introduction}

The physical motivation
of the present paper is that even for quantum description of the
simplest nanoelectronic devices (like a quantum wire) one have to
take into account the interaction of the propagating electron with
the internal degrees of freedom of the device. This interaction
becomes valuable in the case for instance, when the voltage applied
is large enough or the temperature of the device is high. The
simplest excitations in both cases are phonons. It means that the
first step in the correct description of any nanoelectronic device at
nonzero temperature or at large voltage applied is taking into account
the electron phonon-interaction. This leads to polarons as charge
carriers instead of electrons. In general the effective mass of
polaron is greater than the one of electron in the lattice. So one of
the important problem is to calculate the effective mass of charge
carriers and to estimate its contribution into the physical
characteristics of the working regime of the device.

The standard field theoretical
models the electron-phonon interaction such as Fr\"olich model
\cite{Fr} do not preserve the number of particles and cannot be
solved explicitly. This makes interesting the construction of the
phenomenological models of electron-phonon interaction with fixed
number of particles. The first step in this direction is the study
of two-particle models (one electron interacts with one phonon).

The first attempt to construct such a model was undertaken in the
paper \cite{P1} where the electron was described by the
one-dimensional Schr\"odinger operator with the Kr\"onig-Penny
potential and phonon was described by the Jacobi matrix. The
interaction in this model is localized on the diagonal of
electron-phonon space. Being exactly solvable this model however is
not able to calculate explicitly the effective masses of the
quasiparticle aroused. The physical difficulty related to this model
is the difference of scales introduced for description of phonon and
electron (continuous configuration space for electron and the lattice
for phonon).

In the present paper we describe and solve a similar model where the
kinetic energy operators are given by Jacobi matrices for both
electron and phonon. This makes possible to
calculate explicitly the eigenfunctions of the polaron branch of
spectrum and corresponding effective masses.  Another simplification
in comparison to the model from \cite{P1} is that we take into
account the influence of the lattice potential on the electron by
changing of its mass by the effective one in the kinetic energy
operator.

\section{Model}

We start with the one-dimensional lattice ${\bf Z}$.
Set $ H = l^2 ({\bf Z}) $ be the state space of a quantum particle on
the lattice and $T$ be the unitary shift operator in $H$,
$(T\varphi)_n = \varphi_{n-1}$.  Introduce the unperturbed phonon and
electron hamiltonians on $H$ as $ L_1 = T + T^* $ and $ L_2 = \xi (T
+ T^*) $ respectively. Here the real parameter $ \xi $ makes sense
of ratio $\xi \sim \frac {\hbar \omega_0} {E_{char}} $ up to a
dimensionless factor where $\omega_0 $ is the frequency of the
lattice oscillations and $E_{char}$ is a characteristic scale of the
electron energy i.e. the width of conductivity zone.
The hamiltonian of electron-phonon system without interaction acts in
${\cal H} = H \otimes H = l^2 ({\bf Z}^2) $ as $ L_0 = L_1 \otimes I
+ I \otimes L_2 $. Its spectrum is given by $\sigma (L_0) =
\overline{\sigma (L_1) + \sigma (L_2)} = [ -2(1+|\xi |), 2(1+|\xi | )
] $.

We will suppose that the lattice is homogeneous and the interaction
between electron and phonon is localized. This restricts the choice
of the interaction operator $V$, $ L =
L_0 + V $ being the perturbed hamiltonian by the following
assumptions.

{\bf Assumption 1. }.

$[ V, R ] = 0 $, where $ (R g)_{m,n} =
g_{m+1, n+1} $ is the operator of the electron-phonon shift in $\cH
$, - space homogeneity.

{\bf Assumption 2}.

$\forall n \in {\bf Z} \,\,\, g_{n,n} = 0 $ implies $ V g = 0 $, -
locality.

The locality assumption means that $V$ differs from
zero on the main diagonal ${\cal D} = \{ g \in l^2 ({\bf Z}^2):\;
g_{mn} = 0\mbox{ for } m\neq n \} $ of $l^2 ({\bf Z}^2) $ only. By
the assumption ${\bf 1}$ it commutes on ${\cal D}$ with the shift
operator. Thus $V$
is given just by convolution with a function $\Psi \in l^\infty({\bf
Z}) $, $$ (Vg)_{n,m} = \delta_{n,m} \sum_{\tau \in {\bf Z}} \Psi_\tau
g_{n-\tau, n - \tau }. $$

Assuming the symmetricity of $V$ one gets $\Psi_\tau = \overline
{\Psi_{-\tau}} $. If $\Psi_\tau $ is well-behaved (for
instance if $\sum_\tau | \Psi_\tau | < \infty $ and so $V$ is bounded)
then $V$ is selfadjoint. Generally one can define the selfadjoint
operator $V$ in ${\cal D} \simeq l^2 ({\bf Z}) $ as the Fourier
transform of the operator of multiplication on real valued function $
\Psi (p) = \sum_\tau \Psi_\tau e^{ip\tau } $ in $ L^2 [-\pi, \pi ]$.

\subsection{Spectrum of the Operator $L$}

To study operator $ L $ it is convenient to use the
spectral representation of $R$ given by the Gelfand transform
$\cF:\; l^2(\Z^2 )\to L^2([-\pi,\pi], l^2(\Z))\equiv \tcH $, $$
\left(\cF g\right)_s (\kappa ) = \frac 1{\sqrt{2\pi}} \sum_n g_{s+n,n}
e^{i\kappa n}. $$

{\it Properties of $ \cF $}

$ 1^\circ $. $\cF $ is isometric surjection.

{\it Proof }

Set $ D_s = \{ g_{n,m} \neq 0 \Longrightarrow n-m = s\}$ hence $D_s
\perp D_{s^\prime } $ for $ s\neq s^\prime $ and $ \cH = \oplus_{s
\in \Z} D_s $. Denote as $\tD_s = \cF D_s \equiv L^2 (-\pi, \pi )
e_s$, $e_s $ being the $s$-th element of the standard basis in $l^2
(\Z )$. We see that $\tD_s \perp \tD_{s^\prime }$ for $s \neq
s^\prime $ and $\tcH = \oplus \tD_s $. Then $\cF|_{D_s} $ is just
usual Fourier transform. Thus $\cF|_{D_s} $ is surjective isometry.
$\bullet $

$2^\circ $. $\cF $ diagonalizes $ R $, $ R = \cF^{-1} e^{i\kappa }
\cdot\,\cF $.

$3^\circ $. $L_0 = \cF^{-1} \tL_0 \cF $ where $\tL_0 = \int_{-\pi
}^\pi \oplus \tL_0 (\kappa ) d\kappa $ and $$\tL_0 (\kappa ) = (\xi +
e^{-i\kappa} ) T + (\xi +
e^{i\kappa} ) T^* $$
We remind that $T$ is the shift operator in $ l^2 (\Z ) $.

$4^\circ $. $\cF $ diagonalizes $V$, $ V= \cF^{-1} \tV \cF $, where
$$(\tV g)_s (\kappa ) = \delta_{s,0}\Psi (\kappa )g_s (\kappa ). $$

{\it Proof}

Since $D_s$ are invariant subspaces of $R$ and $\cF $ the property
$2^\circ $ expresses the fact that shift turns to multiplication on
function $e^{i\kappa} $ under the Fourier transform. This together
with commutation of $\cF $ with the shift on $s$-variable implies
$3^\circ $. The property $4^\circ $ expresses the fact that under
the Fourier transform the convolution becomes operator of
multiplication on function. $\bullet $

Thus $\tL = \tL_0 + \tV $ is partially diagonalized by $\cF $ i.e.
$$\tL = \int_{-\pi
}^\pi \oplus \tL (\kappa ) d\kappa, $$
$$\tL (\kappa ) = (\xi + e^{-i\kappa} ) T + (\xi +
e^{i\kappa} ) T^* + \Psi (\kappa )P_0 $$
where $(P_0 \varphi)_s = \delta_{s,0}\varphi_s $ is the projection in
the fibre $l^2 ({\bf Z}) $.

Since $\tL( \kappa ) $ is the rank one perturbation of the standard
Jacobi matrix we have $\sigma_c (L_\kappa ) = [ -2 \left|\xi +
e^{i\kappa}\right| , 2 \left|\xi +
e^{i\kappa}\right| ]$. This spectrum corresponds to scattered waves
type eigenfunctions of $L$. It fills the zone $$\sigma_{sc. \,wave}=
\bigcup_{\kappa \in [-\pi,\pi ]}\sigma_c \left(\tL (\kappa )\right) =
[ -2 \left(1 + |\xi |\right) , 2 \left(1+ |\xi |\right) ] $$ which
coincides with the spectrum of unperturbed operator $L_0 $.

\subsection{Polaron}

We are now interested in the waveguide type eigenfunctions of
continuous spectrum of $L$, i.e. those which satisfy the condition $
\varphi_{n,m} = O \left( \exp (- C|n-m|)\right) $. In physical
literature such eigenfunctions are called polarons. It follows from
the previous section that polarons with definite
lengthwise momentum $\kappa \in [-\pi , \pi ] $ i.e. satisfying the
condition $ R\varphi = \chi \varphi $, $\chi = e^{i\kappa}$
correspond to $l^2$-eigenfunctions of $\tL (\kappa ) $.
This reduces the spectral problem $ \tL (\kappa )\varphi = \lambda
\varphi $ to the one for Jacobi matrix $$ Q_\chi = \pmatrix{ \cdots &
\cdots & \cdots & \cdots &  & & \cr \cdots & 0 & \xi + \chi & 0 &
\cdots & \cr \cdots & \xi +\overline \chi & \Psi (\kappa ) & \xi +
\chi & 0 & \cdots \cr \cdots & 0 & \xi + \overline\chi & 0 & \xi +
\chi & 0 \cr & \cdots & 0 & \xi + \overline\chi & 0 &  \xi + \chi \cr
& & \cdots & 0 & \xi + \overline\chi  & 0 \cr & & & \cdots & 0 &
\cdots & \cr} $$ or $$ (Q_\chi \varphi )_n = (\xi +\overline\chi
)\varphi_{n-1} + (\xi + \chi )\varphi_{n+1},\; n \neq 0 $$
$$(Q_\chi \varphi )_0 = (\xi + \overline \chi )\varphi_{-1} +
\Psi (\kappa) \varphi_0 + (\xi + \chi )\varphi_1.  $$ Note that
$$ U^*_\kappa \tL (\kappa ) U _\kappa = |\xi + \chi |\left( J +
\hat{V}\right) $$ where $$ U_\kappa = \mbox {diag }\left\{ e^{i
n\tau}\right\}_{n \in \Z}, \hskip 10mm \tau = \arg (\xi + \chi ),
\hskip 5mm  J_{i,j} = \delta_{i,j+1} + \delta_{i,j-1} $$ and
$$\hat{V} = \frac{\Psi (\kappa ) }{ |\xi + \chi |} P_0 .$$ Thus the
$l^2$-eigenfunction of $Q_\chi $ corresponds to the eigenvalue
$\lambda = | \xi + \chi |\left( a + a^{-1} \right) $ given by
solution of the secular equation $$ a^2 + \frac { \Psi (\kappa )
}{ |\xi + \chi |} a - 1 = 0 \eqno (1)$$ with $|a| < 1 $. Solving this
equation one gets that the polaron energy $\lambda $ is related to
$\kappa $ through the dispersion relation $$\lambda (\kappa ) = \mbox
{sign}\Psi (\kappa) \,\,\sqrt{\Psi (\kappa)^2 + 4 ( 1 + \xi^2 + 2 \xi
\cos \kappa ) }.  \eqno (2) $$  The corresponding eigenfunction of
$\tL (\kappa ) $ has the form $$\varphi_n(\kappa, \lambda) = e^{ i
n\tau} a^{|n|}.$$ Note that equation $(2)$ shows that in the case of
strong coupling ($\Psi (\kappa ) \gg 1 $) the polaron zone is
separated from spectrum corresponding to the scattered waves. For the
small coupling $\Psi (\kappa ) \sim 0 $ the polaron zone and
$\sigma_{sc.  \,wave}$ are overlapped.  Note however that the polaron
decay is impossible for the energies belonging to the common part of
spectrum. This is the result of high common symmetry of $L$ and $L_0$
provided by $R$.

For completeness we now right down the generalized eigenfunctions of
$\tL (\kappa) $ scattered waves type. The solution of the equation
$\tL (\kappa)\varphi =\lambda \varphi $ corresponding to $\lambda =
2|\xi + \chi |\cos \theta $ is given by $$\varphi_n = e^{i n\tau
}\left[ e^{i\theta |n|} + S_\kappa(\lambda) e^{-i \theta |n|}\right]
$$ where $$ S_\kappa(\lambda) = -\frac{\lambda - \Psi (\kappa ) - 2
|\xi + \chi |e^{i\theta }}{\lambda - \Psi (\kappa) - 2 |\xi + \chi |
e^{-i\theta }}. $$ Here $\theta \in [-\pi, \pi ]$ is the
quasimomentum corresponding to direction transverse to the diagonal.
The reflection coefficient $S_\kappa $ has the pole on the physical
sheet in $\lambda $-variable which corresponds to eigenvalue $(2)$.
Since $Q_\chi $ is the rank $1$ perturbation of standard Jacobi
matrix this pole is the only singularity of $S_\kappa $ on the
Riemann surface of the energy $\lambda $.

\subsection{Effective Masses and Fit of the Parameters}

Using $(2)$ one can now
calculate explicitly the effective masses at the ends of the polaron
zone by the formula $ m = \left.\frac 1{\lambda''(\kappa
)}\right|_{\kappa = \kappa_0} $ provided $\lambda'(\kappa_0 ) = 0 $.
The result is $$ m = \left.\frac {\sqrt{\Psi (\kappa)^2 + 4 ( 1 + \xi^2 +
2 \xi \cos \kappa ) }}{{\Psi'}^2 + \Psi \Psi'' - 4\xi \cos \kappa }
\right|_{\kappa = \kappa_0 } $$ where $\kappa_0 $ is defined by
the equation $$\Psi (\kappa_0) \Psi'(\kappa_0 ) = 4 \xi \sin \kappa_0
.$$

The case $ \Psi (\kappa ) =
const $ or $ \Psi_\tau = \psi \delta_{\tau,0} $ comprises a precise
discrete analog of the interaction suggested in the paper \cite{P1}.
The corresponding polaron zone is $\sigma_{pol} = \left[\sqrt{\psi^2
+ (|\xi |- 1)^2},\right. $ $\left.\sqrt{\psi^2 + (|\xi | + 1)^2}
\right] $ for $\psi > 0 $ and $ -\sigma_{pol}$ for $\psi < 0$. The
coupling constant $\psi $ here is an analog of the Fr\"olich constant
$\alpha $ \cite{Fr}. This constant is incorporated in the Fr\"olich
hamiltonian $H$ as $ H = H_0 + \sqrt{\alpha } V $, where $H_0$ is the
operator of kinetic energy of electron-phonon system and $ V$ is the
operator of electron-phonon interaction. In the limit of weak
coupling $\psi \sim 0 $ by $(2)$ we have $\lambda -\lambda_0 \sim
\frac 1{4|\chi + \xi |^2} \psi^2 $, where $ \lambda_0 = 2|\chi + \xi
| $ is the rest of the zone corresponding to scattered waves.
Comparing this with the asymptotic $\Delta \lambda \sim \alpha $ of
the polaron energy measured from the rest of the continuous spectrum
known from physical literature \cite{Pol} one takes $\psi = \tilde{C}
\sqrt{\alpha} $, $\tilde{C} = 2 |\chi + \xi | $.

Then we have $\sin \kappa_0 = 0$ what gives $ \kappa_0 = \pi $
for minimum. Thus $$ m = \frac {\sqrt{\psi^2 + 4 ( 1 - \xi )^2 }}{4
\xi }. \eqno (4)$$
At $\psi \sim 0 $ the polaron effective mass $(4)$ should coincide
with the electron effective mass which is $1$ in the accepted system
of units. This condition fixes the parameter $\xi $ as $\xi = \frac
13$. This defines completely $\tilde{C}$ at the rests of the polaron
zone.

Let us note that the Fr\"olich constant $\alpha $ can be expressed in
terms of the physical characteristics of the material \cite{Fe}
$$\alpha = \frac 12 \left(\varepsilon_\infty^{-1} -
\varepsilon_0^{-1} \right)\frac {e^2}{\hbar \omega_0 }\left(\frac {2
m_e^* \omega }\hbar \right)^{ \frac 12} $$
where $\varepsilon_0 $ and $\varepsilon_\infty  $ are dielectric
constants at zero and high frequencies, $m_e^* $ is the electron
effective mass. Thus we conclude that the coupling constant $\psi $
is completely fixed by the choice of the material. As both of the
parameters $\psi $ and $\xi $ are fitted one can calculate the
polaron effective mass $(4)$ for various materials.

In the weak coupling limit $\psi \sim 0 $ electrons practically
do not interact with
the lattice phonons. Vice versa, in the strong coupling limit
$\psi \to \infty $ electrons strongly interact with the lattice
phonons. In this case the description of the electron-phonon
interaction in the frame of field theory becomes essentially
nonlinear. This means that the linear approximation we used
is not suitable in this regime. That is the reason why the
asymptotics $\lambda \sim \sqrt{\alpha}$, $m \sim \sqrt{\alpha} $ at
$\psi \to \infty $ obtained from $(2)$ under the accepted
fit of $\psi $ differ from the physically expected $\lambda \sim
\alpha^2 $, $ m \sim \alpha^4 $ \cite{LR}.

Note that when $ \xi = 0 $ the polaron zone degenerates
into the point $\lambda = \sqrt {\psi^2 + 4} $. This point is
the infinitely degenerated bound state of $L$. One can interpret
this as the result of braking the electron by vacuum of infinitely
heavy phonons.

\section{Conclusion}

The suggested model of electron-phonon interaction demonstrates on
qualitative level the rising of polaron branch in the spectrum of the
Hamiltonian under consideration while the spectrum of scattered waves
remains the same as for unperturbed problem. This model leads
to the explicit formula for the polaron effective mass. The
latter makes possible to fix the model parameters $\psi $ and $\xi $
as it is described above. Being applied to the description of the
electron propagation through a quantum wire in non-ballistic regime
the model allows to calculate the effective mass of charge carriers
in terms of $\psi $ and $ \xi $. It opens the way for estimation of
the contribution of electron-phonon interaction into the
conductivity in the working regime of any nanoelectronic device
based on quantum wires. The simplest such device is the quantum
wire with the periodic external voltage applied \cite{Pom, IPY, 99, 15,
16, 14, 15, 16,18, a, b, c}. As it is has been shown
\cite{Pom, IPY, 99, 15, 16} in the effective mass of charge carriers is important
characteristic of devices based on Mott-Peierls stimulated transition
and could change the upper limits for the temperature and gate
voltage applied.

\end{document}